%% file: main.tex
\documentclass[3p,times,twocolumn]{elsarticle}

\usepackage{ecrc}
\usepackage{hyperref}

\volume{00}

\firstpage{1}

\journalname{Nuclear Physics B Proceedings Supplement}

\runauth{W. J. Torres Bobadilla et al.}

\jid{nuphbp}

\jnltitlelogo{Nuclear Physics B Proceedings Supplement}

\usepackage{amssymb}

 \biboptions{sort&compress}

\usepackage[figuresright]{rotating}
\usepackage{graphicx}  
\usepackage{dcolumn}   
\usepackage{bm}        
\usepackage{amssymb}   
\usepackage{amsmath}
\usepackage{slashed}
\usepackage{stackrel}
\usepackage{feynarts}
\usepackage{color}

\RequirePackage{flushend}

\usepackage{multirow}
\usepackage{array}

\allowdisplaybreaks

\newcommand{\beq}{\begin{equation}}
\newcommand{\eeq}{\end{equation}}
\newcommand{\bqa}{\begin{eqnarray}}
\newcommand{\eqa}{\end{eqnarray}}

\newcommand{\nn}{\pagebreak[0] \nonumber \\}

\def\sam{{{\sc S@M}}}

\def\njet{{{\sc NJet}}}

\def\C++{{{\sc c++ }}}
\def\Feynarts{{{\sc FeynArts}}}
\def\Feyncalc{{{\sc FeynCalc}}}

\newcommand{\cp}{\mbox{c.p.}}
\newcommand{\QQ}{Q}
\newcommand{\GG}{G}
\newcommand{\GH}{\hat G}
\newcommand{\GA}{\Gamma}

\begin{document}

\begin{frontmatter}



\newcommand{\mpi}{Max-Planck-Institut f\"ur Physik, F\"ohringer Ring 6, 80805 M\"unchen, Germany}
\newcommand{\padova}{Dipartimento di Fisica e Astronomia, Universit\`a di Padova, and INFN  \\   Sezione di Padova, via Marzolo 8, 35131 Padova, Italy}                       
\newcommand{\bogota}{Departamento de F\'isica, Universidad Nacional de Colombia,  Ciudad Universitaria, Bogot\'a, D.C. Colombia}

\dochead{}

\title{Generalised Unitarity for Dimensionally Regulated Amplitudes}

\author[w1]{W. J. Torres Bobadilla\corref{label1}}
\author[w2]{A. R. Fazio}
\author[w1,w3]{P. Mastrolia}
\author[w3]{E. Mirabella}
\address[w1]{\padova}
\address[w2]{\bogota}
\address[w3]{\mpi}
\cortext[label1]{Speaker and Corresponding Author}


\begin{abstract}
We present a novel set of Feynman rules and generalised unitarity
cut-conditions
for computing one-loop amplitudes via $d$-dimensional integrand
reduction algorithm.
Our algorithm is suited for analytic as well as numerical result,
because all ingredients turn out to have a four-dimensional
representation.
We will apply this formalism to NLO QCD corrections.

\end{abstract}

\begin{keyword}
Quantum Chromodynamics\sep Scattering amplitudes\sep Next-to-leading-order\sep Generalised Unitarity.
\end{keyword}

\end{frontmatter}



\input{Section1}
\input{Section2}

\input{Section3}
\input{Section4}
\input{Section6}

\section*{Acknowledgements}
We would like to thank Valery Yundin for comparison of gluon amplitudes with \njet\, and Tiziano Peraro for useful discussions. A.R.F. is partially supported by the UNAL-DIB grant N. 20629 of  the ``Convocatoria del programa nacional de proyectos para el fortalecimiento de la investigaci\'on, la creaci\'on y la innovaci\'on en posgrados de la Universidad  Nacional de Colombia 2013-2015". \\
The work of   P.M.  is  supported by the Alexander von Humboldt Foundation, in the framework of the Sofja Kovaleskaja Award Project
``Advanced Mathematical Methods for Particle Physics'', endowed by the German Federal Ministry of Education and Research.  \\
W.J.T. is supported by Fondazione Cassa di Risparmio di Padova e Rovigo (CARIPARO).




\bibliographystyle{elsarticle-num}
\bibliography{references}

\end{document}

%% file: Section1.tex
\section{Introduction}
\label{sec:intro}
In the era of the LHC experiments of increasing accuracy become possible, where one of the highlight of Run 1 of the LHC was the discovery by CMS and ATLAS of a Higgs boson~\cite{Aad:2012tfa,Chatrchyan:2012ufa}. Hence it is necessary to achieve more accurate results for measurable quantities at the theoretical level.\\
According to perturbation theory, higher order corrections to amplitudes have to be considered. To evaluate such corrections in quantum field theory, it is necessary to compute multi loop Feynman diagrams, where, instead of the explicit set of loop Feynman diagrams, the basic reference point is the linear expansion of the amplitude function in a basis of master integrals (MI's), multiplied by coefficients that are rational functions of the kinematic variables, already known as Passarino Veltman reduction theorem~\cite{Passarino:1978jh} at the level of one-loop.\\
It is in fact possible to recover the finiteness of scattering amplitudes
at integrand level by constructing the integrands by a multi-particle
pole expansion arising from the analiticity properties and unitarity
of the S-matrix. Indeed, scattering amplitudes, continued for complex
momenta, reveal their singularities structures as poles and branch
cuts. The unitarity based method (UBM) allows to determine the coefficients of the MI's by
expanding the integrand of the tree level cut amplitudes into an expression
that resembles the cut of the basis integrals.

In this talk, I review the four dimensional formulation (FDF) proposed in~\cite{Fazio:2014xea} which is equivalent to the four-dimensional helicity (FDH) scheme~\cite{Bern:1991aq,Bern:1995db,Bern:2002zk}, and 
allows for a purely four-dimensional regularisation of the amplitudes.
Within FDF, the states in the loop are described as 
four dimensional massive particles. The four-dimensional degrees
of freedom of the gauge bosons are carried by  {\it massive vector bosons}
of mass $\mu$ and their 
 $(d-4)$-dimensional ones by {\it real scalar particles} obeying 
a simple set of four-dimensional Feynman rules. A $d$-dimensional fermion 
of mass $m$ is instead traded for a {\it tardyonic  Dirac field} with
mass $m +i \mu \gamma^5$~\cite{Jentschura:2012vp}.   The $d$ dimensional algebraic manipulations are 
replaced by four-dimensional ones complemented by    
a set of multiplicative selection rules. The latter are treated as an algebra
describing internal symmetries.  

This contribution is organised  as follows:  section~\ref{sec:4D} is devoted to the description
of the regularisation method, while  Section~\ref{sec:onshell} describes how 
generalised  unitarity method can be  applied in presence of a  FDF 
of one-loop amplitudes. Section~\ref{sec:4P} shows 
the decomposition in terms of MIs 
of certain classes of $2 \to (n-2)$ one-loop amplitudes. Section~\ref{examples} collects the applications of generalised unitarity methods within the FDF.  In particular it presents results
for representative helicity amplitudes of  $gg\to gg$, $gg\to ggg$, $gg\to gggg$ and $gg\to g\text{H}$.

%% file: Section2.tex
\section{Four dimensional formulation of the $d$-dimensional regularisation scheme}
\label{sec:4D}
We discuss briefly the regularisation scheme proposed in~\cite{Fazio:2014xea}, only pointing out the main ingredients.
\par Let's denote as barred a quantity referred to unobserved particles and living therefore in a $d_s$-dimensional space. Then, the metric tensor can be split as
\begin{align}
\bar g^{\mu \nu} = g^{\mu \nu} + \tilde g^{\mu \nu}  \, ,
\end{align}
in terms of  a four-dimensional tensor  $g$ and a $-2 \epsilon$-dimensional one, $ \tilde g$,
such that
\begin{align}
 \tilde g^{\mu \rho} \, g_{\rho \nu} =0 \, , \qquad 
 \tilde g^{\mu}_{\phantom{\mu}  \mu}  =  -2 \epsilon \stackrel[d_s\to4]{}{\longrightarrow} 0 \, , \qquad 
 g^{\mu}_{\phantom{\mu} \mu}  = 4 \, , 
 \label{Eq:OrthoGs}
\end{align}  
The  tensors $g$ and $\tilde g$  project a $d_s$-dimensional vector $\bar q$ into the four-dimensional  and the  
$-2 \epsilon$-dimensional subspaces respectively, 
\begin{align}
q^\mu &\equiv  g^{\mu}_{\phantom{\mu} \nu} \,  \bar q^\nu\, ,  &  \tilde q^\mu &\equiv \tilde g^{\mu}_{\phantom{\mu} \nu} \, \bar q^\nu\, .
\label{Eq:Projection}
\end{align}
and the properties for the matrices $\tilde \gamma^\mu = \tilde g^{\mu}_{\phantom{\mu} \nu} \, \bar \gamma^\nu$ can be obtained from Eq.~(\ref{Eq:OrthoGs})
\begin{subequations}
\begin{align}
[ \tilde \gamma^{\alpha}, \gamma^{5}   ] &= 0 \, , & 
\{\tilde \gamma^{\alpha}, \gamma^{\mu}  \} &=0 \ , \label{Eq:Gamma01} \\
\{\tilde \gamma^{\alpha}, \tilde \gamma^{\beta}  \} &= 2 \,  \tilde g^{\alpha \beta} \, . 
\end{align}\label{Eq:Gamma02}
\end{subequations}
In principle, we could infer the behaviour of $\tilde \gamma^{\alpha}$ from (\ref{Eq:Gamma02}) and say $\tilde \gamma \sim \gamma^5$, however, this choice does not fulfil the Clifford algebra when $d_s\to4$. It means we cannot have any four-dimensional representation of the $-2\epsilon$-subspace, therefore, we introduce an algebra with an internal symmetry called $-2\epsilon$ \textit{selection rules}, $(-2\epsilon)$-SRs, which consists in performing the substitutions
\begin{align}
\tilde g^{\alpha \beta} \to   G^{AB}, \qquad  \tilde \ell^{\alpha} \to i \, \mu \, \QQ^A \; , \qquad  \tilde \gamma^\alpha \to \gamma^5 \, \GA^A\, .
\label{Eq:SubF}
\end{align} 
The $-2\epsilon$-dimensional vectorial
 indices  are thus  traded for  ($-2\epsilon$)-SRs  such that
\begin{align}
\GG^{AB}\GG^{BC} &= \GG^{AC},    & \GG^{AA}&=0,  &   \GG^{AB}&=\GG^{BA},  \nn
 \GA^A \GG^{AB} &= \GA^B,              &   \GA^A \GA^{A} &=0,  & \QQ^A \GA^{A}  &=1, \nn
\QQ^A \GG^{AB} &= \QQ^B,              & \QQ^A \QQ^{A} &=1.  
\label{Eq:2epsA}
\end{align}
The exclusion  of the terms containing odd powers of $\mu$ completely
defines the FDF, and allows one to build integrands which, upon
integration, yield to the same result as in the FDH scheme.

%% file: Section3.tex
\section{Generalised Unitarity}
\label{sec:onshell}
In this section we discuss the consequence of using internal lines in $(4-2\epsilon)$-dimensions within FDF where spinors and polarisation vectors are written explicitly. These ingredients allow the construction of the tree-level amplitudes that are needed to recover any one-loop amplitude.\\
Due to FDF scheme is suitable for the four-dimensional formulation of $d$-dimensional 
generalised unitarity, all kinematics in the construction of the amplitude admit an explicit representation in terms of generalised spinors and polarisation expressions.
\smallskip

In the following discussion we will decompose a   $d$-dimensional  momentum $\bar \ell$ as follows
\begin{align}
\bar \ell = \ell + \tilde \ell \, , \qquad \bar \ell^2  = \ell^2 -\mu^2 = m^2  \,  ,
\label{Eq:Dec0}
\end{align}
while its four-dimensional component $\ell$ will be expressed as 
\begin{align}
\ell = \ell^\flat + \hat q_\ell \, , \qquad \hat q_\ell \equiv \frac{m^2+\mu^2 }{2\, \ell \cdot q_\ell} q_\ell  \,  ,
\label{Eq:Dec}
\end{align}
in terms of the two massless  momenta $\ell^\flat$ and $q_\ell$.

\subsection{Spinors}
The spinors of a $d$-dimensional fermion have to fulfil a completeness relation which
reconstructs the numerator of the cut  propagator,
\begin{small}
\begin{align}
\sum_{\lambda=1}^{2^{(d_s-2)/2}}  u_{\lambda, \, (d)}\left(\bar \ell \right)\bar{u}_{\lambda, \, (d)}\left(\bar \ell \right) & =\bar{ \slashed \ell}+m \, , \nn
\sum_{\lambda=1}^{2^{(d_s-2)/2}}  v_{\lambda, \, (d)}\left(\bar \ell \right)\bar{v}_{\lambda, \, (d)}\left(\bar \ell \right) & =\bar{ \slashed \ell}-m \, .
\label{Eq:CompFD}
\end{align}
\end{small}
The  substitutions~(\ref{Eq:SubF}) allow one to express  Eq.~(\ref{Eq:CompFD}) as follows:
\begin{align}
\sum_{\lambda=\pm}u_{\lambda}\left(\ell \right)\bar{u}_{\lambda}\left(\ell \right) & = \slashed \ell + i \mu \gamma^5 + m \, , \nn
\sum_{\lambda=\pm}v_{\lambda}\left(\ell  \right)\bar{v}_{\lambda}\left(\ell \right)  & = \slashed \ell + i \mu \gamma^5  - m \, .
\label{Eq:CompF4}
\end{align}
with the generalised massive spinors~\cite{Fazio:2014xea}
\begin{subequations}
\begin{small}
\begin{align}
u_{+}\left(\ell \right) & =\left| \ell^{\flat}\right\rangle +\frac{\left(m - i\mu\right)}{\left[ \ell^{\flat}\,q_\ell \right]}\left|q_\ell \right]\, ,  &
u_{-}\left(\ell \right) & =\left| \ell^{\flat}\right]+\frac{\left(m  +  i\mu\right)}{\left\langle \ell^{\flat}\, q_\ell \right\rangle }\left|q_\ell \right\rangle , \nn
v_{-}\left(\ell \right) & =\left| \ell^{\flat}\right\rangle -\frac{\left(m  -  i\mu\right)}{\left[ \ell^{\flat}\, q_\ell \right]}\left|q_\ell \right]\, ,  &
\label{Eq:SpinorG}
v_{+}\left(\ell \right) & =\left| \ell^{\flat}\right]-\frac{\left(m +  i\mu\right)}{\left\langle \ell^{\flat}\, q_\ell \right\rangle }\left|q_\ell \right\rangle , \\[3ex]
\bar{u}_{+}\left(\ell \right) & =\left[\ell^{\flat}\right|+\frac{\left(m + i\mu\right)}{\left\langle q_\ell\, \ell^{\flat}\right\rangle }\left\langle q_\ell\right|\, , &
\bar{u}_{-}\left(\ell \right) & =\left\langle \ell^{\flat}\right|+\frac{\left(m -  i\mu\right)}{\left[q_\ell\,  \ell^{\flat}\right]}\left[q_\ell\right| \, , \nn
\bar{v}_{-}\left(\ell \right) & =\left[\ell^{\flat}\right|-\frac{\left(m  +  i\mu\right)}{\left\langle q_\ell  \, \ell^{\flat}\right\rangle }\left\langle q_\ell \right| \, ,  &
\bar{v}_{+}\left(\ell \right) & =\left\langle \ell^{\flat}\right|-\frac{\left(m - i\mu\right)}{\left[q_\ell\,  \ell^{\flat}\right]}\left[q_\ell \right| \, ,
\end{align}
\end{small}
\label{Eq:SpinorF}
\end{subequations}
\noindent fulfil the completeness relation~(\ref{Eq:CompF4}).

\subsection{Polarisation vectors}
In the axial gauge, the helicity sum of the transverse polarisation vector is
\begin{small}
\begin{equation}
\sum_{i=1}^{d-2}\,\varepsilon_{i\,(d)}^{\mu}\left(\bar{\ell},\bar{\eta}\right)\varepsilon_{i\,(d)}^{\ast\nu}\left(\bar{\ell},\bar{\eta}\right)=-\bar{g}^{\mu\nu}+\frac{\bar{\ell}^{\mu}\,\bar{\eta}^{\nu}+\bar{\ell}^{\nu}\,\bar{\eta}^{\mu}}{\bar{\ell}\cdot\bar{\eta}}\,,\label{Eq:CompGD}
\end{equation}
\end{small}
where $\bar \eta$ is an arbitrary $d$-dimensional massless momentum such that $\bar \ell \cdot \bar \eta \neq 0$.\\
In particular the choice
\begin{equation}
\bar \eta^\mu = \ell^\mu - \tilde \ell^\mu \, ,
\end{equation}
with $\ell$, $\tilde \ell$ defined in Eq.~(\ref{Eq:Dec0}), 
allows us to disentangle the four-dimensional contribution form the $d$-dimensional one:  
\begin{small}
\begin{align}
\sum_{i=1}^{d -2} \, \varepsilon_{i\, (d)}^\mu\left (\bar \ell , \bar \eta \right )\varepsilon_{i\, (d)}^{\ast \nu}\left (\bar \ell , \bar \eta \right ) &=\left (   - g^{\mu \nu}  +\frac{ \ell^\mu \ell^\nu}{\mu^2} \right) -\left (  \tilde g^{\mu \nu}  +
\frac{ \tilde \ell^\mu \tilde \ell^\nu}{\mu^2} \right ) \, .
\label{Eq:CompGD2}
\end{align}
\end{small}
The first term is  related to the cut propagator of a massive gluon  and can be expressed as follows
\begin{align}
 -g^{\mu\nu}+\frac{\ell^{\mu}\ell^{\nu}}{\mu^{2}} &=  \sum_{\lambda=\pm,0}\varepsilon_{\lambda}^{\mu}(\ell) \, \varepsilon_{\lambda}^{*\nu}(\ell)  \, ,  \label{flat}
 \end{align}
in terms of the polarisation vectors of a vector boson of mass $\mu$~\cite{Mahlon:1998jd},  
\begin{align}
&\varepsilon_{+}^{\mu}\left(\ell \right)  = -\frac{\left[\ell^{\flat}\left|\gamma^{\mu}\right|  \hat q_\ell \right\rangle }{\sqrt{2}\mu},
&\varepsilon_{-}^{\mu}\left(\ell \right)     = - \frac{\left\langle \ell^{\flat}\left|\gamma^{\mu}\right| \hat q_\ell \right]}{\sqrt{2}\mu},
\notag\\  
&\varepsilon_{0}^{\mu}\left(\ell   \right)  =   \frac{\ell^{\flat\mu}-\hat q_\ell^{\mu}}{\mu} \, .
\label{emass1}
\end{align}
These polarisation vectors are orthonormal and display all of the usual properties expected for massive vector bosons
 \begin{align}
 \varepsilon^2_{\pm}(\ell) & =\phantom{-} 0\, ,  & \varepsilon_{\pm}(\ell)\cdot\varepsilon_{\mp}(\ell)&=-1\, , \nn
 \varepsilon_{0}^2(\ell)& =-1\, , &  \varepsilon_{\pm}(\ell)\cdot\varepsilon_{0}(\ell)   &=\phantom{-} 0\, ,  \nn
  \varepsilon_{\lambda}(\ell) \cdot \ell &=\phantom{-} 0 \, .
  \label{Eq:propEps}
 \end{align}
The second term of the r.h.s. of Eq.~(\ref{Eq:CompGD2})  is related to the numerator of cut propagator of the scalar $s_g$ and can be expressed in terms
 of the $(-2 \epsilon)$-SRs as:
 \begin{equation}
 \tilde g^{\mu \nu}  +\frac{ \tilde \ell^\mu \tilde \ell^\nu}{\mu^2}  \quad \to  \quad  \GH^{AB} \equiv  \GG^{AB} - \QQ^A \QQ^B  \, .
 \label{Eq:Pref}
\end{equation}
The factor $\GH^{AB}$ can be easily accounted by defining the cut propagator as
\vspace{-0.8cm}
\begin{equation}
\parbox{25mm}{\input{FeynmanRules/SS1.tex}} =  \hat G^{AB}\,\delta^{ab} \, .
\vspace{-0.6cm} 
\label{Eq:Rules2a}
\end{equation}
From generalised spinors and polarisation vectors the $\mu$-dependence of the tree-level amplitude arises.\\
The FDF approach to reconstruct the rational part of one-loop scattering amplitudes is different from the supersymmetric decomposition~\cite{Brandhuber:2005jw} and from the six-dimensional formalism~\cite{Davies:2011vt}. Indeed, to compute any one-loop amplitude via supersymmetric decomposition one splits the amplitude in two terms: \textit{i}) cut constructible part which is obtained by using four-dimensional unitarity, \textit{ii}) and the rational one that is reached by introducing in the amplitude a complex scalar in $d$-dimensions and deal with a massive four-dimensional ones.\\
On the other hand, the six-dimensional helicity method treats $d$-dimensional on-shell momenta into a six-dimensional massless basis and, on the cuts, uses six dimensional helicity spinors to compute the relevant tree-level amplitudes. However, because of the argument given in~\cite{Giele:2008ve}, the contribution that comes from this treatment gives a result that has to be corrected by hand with the help of topologies involving complex scalars along the lines.\\
Unlike the approaches presented above, FDF does not make any distinction between cut-constructible or rational part, as well, the result obtained with FDF scheme is automated corrected by the $(-2\epsilon)$-SRs, it splits the $d$-dimensional objects into their four-dimensional and $(d-4)$-dimensional parts and finds a four-dimensional representation for both of them. Moreover, the approaches already described are simpler than the ones that introduce explicit higher-dimensional extension of either the Dirac~\cite{Giele:2008ve,Ellis:2008ir} or the spinor~\cite{Cheung:2009dc,Davies:2011vt} algebra.

%% file: FeynmanRules/SS1.tex
\unitlength=0.25bp%
\begin{feynartspicture}(300,300)(1,1)
\FADiagram{}
\FAProp(4.,10.)(16.,10.)(0.,){/ScalarDash}{0}
\FALabel(5.5,8.93)[t]{\tiny $a,A$}
\FALabel(14.5,8.93)[t]{\tiny $b, B$}
\FAVert(4.,10.){0}
\FAVert(16.,10.){0}
\FAProp(10.,6.)(10.,14.)(0.,){/GhostDash}{0}
\end{feynartspicture}

%% file: Section4.tex
\section{One-loop amplitudes}
\label{sec:4P}

In order to apply generalised-unitarity methods within FDF, we consider as examples the one-loop  $2 \to 2,3,4$ scattering amplitudes, where external particles are gluons.
\smallskip

In general, due to the reduction theorem any massless four-point one-loop  amplitude  can be
decomposed in terms MIs, as follows 
\begin{subequations}
\begin{align}
A_n^{\text{1-loop}} &{}= \frac{1}{(4 \pi)^{2-\epsilon}}   \,   \sum_{i<j<k<l}^{n-1} \bigg [ c_{i|j|k|l;\,0}\, I_{i|j|k|l}+c_{ij|k|l;\,0}\, I_{ij|k|l}   \nn
&{} \quad +c_{ij|kl;\,0}\, I_{ij|kl}\bigg ] +  \mathcal{R}\, , \label{Eq:Deco} \\
\mathcal{R} &{}=    \frac{1}{(4 \pi)^{2-\epsilon}} \bigg [ \, c_{i|j|k|l;\,4}\, I_{i|j|k|l}[\mu^{4}]+\big(c_{ij|k|l;\,2}\, I_{ij|k|l}[\mu^{2}]  \nn
&{} \quad+ c_{ij|kl;\,2}\, I_{ij|kl}[\mu^{2}] \bigg ]  \, . \label{Eq:Ratio} 
\end{align}
\label{Eq:Decomposition}
\end{subequations}

In Eq.~(\ref{Eq:Decomposition}), we see the decomposition between cut-constructible and rational part,  where the latter has been collected in ${\cal R}$. However, we emphasise one-loop processes are not computed by distinguishing those two pieces, instead within the FDF the two contributions are computed simultaneously from the same cuts.\\

The coefficients $c$'s entering in the decompositions~(\ref{Eq:Decomposition}) can 
be obtained by using the generalised unitarity techniques  for quadruple~\cite{Britto:2004nc,Badger:2008cm}, 
triple~\cite{Mastrolia:2006ki,Forde:2007mi,Badger:2008cm}, and double~\cite{Britto:2005ha, Britto:2006sj, Mastrolia:2009dr} cuts.  Since internal particles are massless the single-cut 
techniques~\cite{Kilgore:2007qr,Britto:2009wz, Britto:2010um} are not needed.  In general, the cut  $C_{i_1\cdots i_k}$, defined by the conditions $D_{i_1} =\cdots = D_{i_k}=0$,  allows for the determination of the coefficients $c_{i_1\cdots i_k; \, n}$.

\section{Examples}
\label{examples}
\subsection{The all plus four-gluon amplitude}
\label{AllPlus}

First, let us consider one-loop four-point amplitudes 
with four outgoing  massless particles 
\begin{align}
0 \to 1(p_1) \, 2(p_2) \, 3(p_3)  \, 4(p_4) \,  ,
\label{Eq:ProM}
\end{align}
where $p_i$ is the momentum of the particle $i$.
\smallskip

Within the FDF, we consider the colour-ordered Feynman rules that contain interactions between gluons and scalars, however, due to the $(-2\epsilon)$-SRs, the relevant interactions are: \textit{i}) three gluons and \textit{ii}) one gluon with two scalars, see the discussion below.\\
\noindent Let us compute the four-gluon colour-ordered helicity amplitude $A_{4}\left(1_{g}^{+},2_{g}^{+},3_{g}^{+},4_{g}^{+}\right)$, which at tree-level vanishes, while the one-loop contribution is finite and is obtained from the quadruple-cut $C_{1|2|3|4}$.\\
Since contribution to this amplitude comes only from the boxes and in FDF we have five boxes, we  decompose this sum of boxes as:
\begin{align}
C_{1|2|3|4}& = \sum_{i=0}^4 \, C^{[i]}_{1|2|3|4} \, ,  & c_{1|2|3|4;\, n} &= \sum_{i=0}^4 \, c^{[i]}_{1|2|3|4; \,n} \, ,
\end{align}
where  $C^{[i]}$ ($c^{[i]}$) is  the contribution to the cut (coefficient) involving $i$ internal scalars. \\
The quadruple cuts read as follows
\begin{subequations} 
\begin{align}
C^{[0]}_{1|2|3|4} &{}= 
      \parbox{20mm}{\input{DiagramCut/Cg1234.1.tex}}
    \!+\!  \parbox{20mm}{\input{DiagramCut/Cg1234.2.tex}} 
\! +\!  \parbox{20mm}{\input{DiagramCut/Cg1234.3.tex}}  
\, ,  \label{Will4g} \\[-5.ex] 
C^{[1]}_{1|2|3|4} &{}= \sum_{h_i = \pm, 0}  \,  \mathcal{T}_1 \!\!\! 
\parbox{20mm}{\input{DiagramCut/Cs1.tex}}\! +\!  \cp     \, ,  \label{Eq:Scal1} \\[-5.ex]  
C^{[2]}_{1|2|3|4} &{}=  \sum_{h_i = \pm, 0}  \,  \mathcal{T}^2_1 \!\!\! 
\parbox{20mm}{\input{DiagramCut/Cs2o.tex}}  
+   \mathcal{T}_2\!\!\! 
\parbox{20mm}{\input{DiagramCut/Cs2a.tex}}  \!+\! \cp   \, ,  \label{Eq:Scal2}   \\[-5.ex]  
C^{[3]}_{1|2|3|4} &{}=  \sum_{h_1 = \pm, 0}  \,    \mathcal{T}_3 \!\!\!  
\parbox{20mm}{\input{DiagramCut/Cs3.tex}}  \!+\! \cp   \, ,  \label{Eq:Scal3}  \\[-5.ex]
C^{[4]}_{1|2|3|4} &{}=   \mathcal{T}_4   \!\!\! 
\parbox{20mm}{\input{DiagramCut/Cg1234.4.tex}}
\, , 
\label{Eq:AllScalars} 
\end{align}
\label{Eq:AllCut}
\end{subequations}
where  the abbreviation ``c.p."  means ``cyclic permutations of the external particles''.  In Eqs.~(\ref{Eq:AllCut}) , the $(-2\epsilon)$-SR
have been stripped off and collected in the prefactors $\mathcal{T}_i$,
\begin{align}
&\mathcal{T}_1 &{}={}& \QQ^A\GH^{AB} \QQ^B &{}={}& \phantom{-} 0 \,  ,  \nn
&\mathcal{T}_2 &{}={}& \QQ^A \GH^{AB} \GG^{BC}  \GH^{CD} \QQ^D&{}={}& \phantom{-}  0 \,  , \nn
&\mathcal{T}_3 &{}={}& \QQ^A \GH^{AB} \GG^{BC}  \GH^{CD} \GG^{DE} \GH^{EF} \QQ^F&{}={}& \phantom{-} 0  \, , \nn
&\mathcal{T}_4 &{}={}& \mbox{tr} \left (  \GG \,  \GH \,  \GG \,   \GH \,  \GG \,  \GH \,  \GG \,  \GH \right ) &{}={}& - 1 \, .
\end{align}
The prefactors  $\mathcal{T}_1, \ldots , \mathcal{T}_3$ force the cuts~(\ref{Eq:Scal1}) - (\ref{Eq:Scal3}) to vanish identically.
The only cuts contributing, Eqs.~(\ref{Will4g}) and~(\ref{Eq:AllScalars}), lead to the following  coefficients
\begin{align}
c^{[0]}_{1|2|3|4; \; 0} &=0  \, ,
& c^{[0]}_{1|2|3|4; \; 4} &=3 i\frac{\left[12\right]\left[34\right]}{\left\langle 12\right\rangle \left\langle 34\right\rangle } \,  ,   \nn
c^{[4]}_{1|2|3|4; \; 0} &=0 \, ,
& c^{[4]}_{1|2|3|4; \; 4} &= - i\frac{\left[12\right]\left[34\right]}{\left\langle 12\right\rangle \left\langle 34\right\rangle }  \,  .
\end{align}
Therefore the only non-vanishing   coefficient, $c_{1|2|3|4; \; 4}$,    is   
\begin{align}
c_{1|2|3|4; \; 4} &{}= c^{[0]}_{1|2|3|4; \; 4} + c^{[4]}_{1|2|3|4; \; 4} =
2i\frac{\left[12\right]\left[34\right]}{\left\langle 12\right\rangle \left\langle 34\right\rangle } \, .
\end{align}
The colour-ordered one-loop amplitude can be obtained from
Eq.~(\ref{Eq:Decomposition}), which in this simple case reduces to 
\begin{align}
A_{4}\left(1_{g}^{+},2_{g}^{+},3_{g}^{+},4_{g}^{+}\right) &=
c_{1|2|3|4; \; 4} \ I_{1|2|3|4}[\mu^4] \notag\\
&= -\frac{i}{48 \, \pi^2} \, \frac{\left[12\right]\left[34\right]}{\left\langle 12\right\rangle \left\langle 34\right\rangle } \,  ,
\label{Eq:Res4G}
\end{align}
and is in agreement with the literature~\cite{Kunszt:1993sd}. \\
\smallskip
It is worth to mention the $(-2\epsilon)$-SRs can be performed before the tree-level amplitudes are computed as well automated cut-by-cut. It means topologies which vanish because of the $(-2\epsilon)$-SRs can be removed at the beginning of the computation without affecting our result.\\

\par\bigskip In ~\cite{Fazio:2014xea} the method is also checked by computing the QCD one loop correction to the scattering amplitude of two gluons production by quark anti-quark annihilation and the processes at two-loop of three gluons fusion in a Higgs.

\subsection{Five- and six-gluon amplitudes}
In this section we show how FDF works at higher point amplitudes, as an explicit example we present the analytic contribution of the five- and six- point all plus amplitude, we choose this helicity configuration because of the absence of  triangles and bubble contributions. 

\subsubsection{Five gluon amplitudes}
\begin{align}
C_{12|3|4|5} &{} = \quad 
\parbox{20mm}{\input{FeynmanDiagrams/5gQ.tex}}
\quad+\quad\parbox{20mm}{\input{FeynmanDiagrams/5sQ.tex}} \quad , \nn
c_{12|3|4|5;\,0}&{}=0\, , \nn
c_{12|3|4|5;\,4}&{}= \frac{2i\left[21\right]\left[43\right]\left[53\right]\left[54\right]}{\left\langle 12\right\rangle \text{tr}_{5}\left(4,1,5,3\right)}\, ;  
\end{align}
with $ \text{tr}_{5}\left(1,2,3,4\right)=\left\langle 1\left|234\right|1\right]-\left[1\left|234\right|1\right\rangle$.\\
From eq.~(\ref{Eq:Decomposition}), the finite colour-ordered one-loop amplitude reduces to
\begin{align}
A_{5}^{\text{1-loop}}\left(1^{+},2^{+},3^{+},4^{+},5^{+}\right)&=c_{12|3|4|5;\,4}\, I_{12|3|4|5}\left[\mu^{4}\right]\nn
&\qquad+\text{cyclic perms.}
\end{align}
In agreement agrees with~\cite{Bern:1993qk}.\\

\par
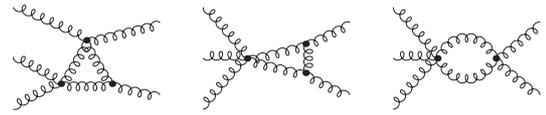
\begin{figure}[htb!]
\begin{subequations}
\begin{align*}
&\parbox{25mm}{\input{FeynmanDiagrams/5gT2.tex}}
\parbox{25mm}{\input{FeynmanDiagrams/5gT3.tex}}
\parbox{25mm}{\input{FeynmanDiagrams/5gD.tex}}
\end{align*}
\end{subequations}
\caption{Triangle and bubble topologies for the five-point.}\label{5top}
\end{figure}

 It is worth to mention that within FDF we have also computed other helicity configurations, 
$A_{5}\left(1^{-},2^{+},3^{+},4^{+},5^{+}\right)$,  $A_{5}\left(1^{-},2^{-},3^{+},4^{+},5^{+}\right)$ and $A_{5}\left(1^{-},2^{+},3^{-},4^{+},5^{+}\right)$, where contributions from triangles and bubbles arise, then, to obtain these contributions we consider the topologies showed in Fig.~\ref{5top}, with the following trees needed as input:
\begin{align}
&G\, G\to g,&&S\, S\to g,\nn
&G\, G\to g\, g,&&S\, S\to g\, g,\nn
&G\, G\to g\, g\, g,&&S\, S\to g\, g\, g.\label{trees5g}
\end{align}

Where $G$ and $S$ are the generalised gluon and colour scalar respectively and, $g$ represents the external gluon. As was discussed in section \ref{AllPlus}, tree levels containing both generalised gluon and colour scalar as internal legs do not contribute to the coefficient due to the $-2\epsilon$-SRs.
\par\noindent It means that to recover any five-gluon amplitude for a particular helicity configuration we compute the coefficients that appear in eq.~(\ref{Eq:Decomposition}), obtaining an agreement with \njet~\cite{Badger:2012pg} and reproducing previous results~\cite{Bern:1993mq}.
\vspace{-0.4cm}

\subsubsection{Six gluon amplitudes}
\begin{subequations}
\begin{align}
C_{123|4|5|6} &{} = \quad 
\parbox{20mm}{\input{FeynmanDiagrams/6gQ1.tex}}
\quad+\quad\parbox{20mm}{\input{FeynmanDiagrams/6sQ1.tex}}\quad , \nn
c_{123|4|5|6;\,0}&{}=0\, , \nn
c_{123|4|5|6;\,4}&{}=\frac{2i\left[56\right]}{\left\langle 12\right\rangle \left\langle 23\right\rangle \text{tr}_{5}\left(5,4,6,1\right)\text{tr}_{5}\left(5,4,6,3\right)}\nn
&\qquad\times\bigg(s_{45}\left\langle 6\left|1+2\right|3\right]\left[51\right]\left[64\right]^{2}\nn
&	\qquad\qquad-s_{46}\left\langle 5\left|1+2\right|3\right]\left[54\right]^{2}\left[61\right]\bigg)\, ;  \\
C_{12|34|5|6} &{} = \quad 
\parbox{20mm}{\input{FeynmanDiagrams/6gQ2.tex}}
\quad+\quad\parbox{20mm}{\input{FeynmanDiagrams/6sQ2.tex}}\quad , \nn
c_{12|34|5|6;\,0}&{}=0\, , \nn
c_{12|34|5|6;\,4}&{}=\frac{2i\left\langle 5\left|1+2\right|6\right]\left\langle 6\left|1+2\right|5\right]\left[12\right]\left[43\right]\left[65\right]^{2}}{\left\langle 12\right\rangle \left\langle 23\right\rangle \text{tr}_{5}\left(5,2,6,1\right)\text{tr}_{5}\left(5,4,6,3\right)} \, ;  \\
C_{12|3|45|6} &{} = \quad 
\parbox{20mm}{\input{FeynmanDiagrams/6gQ3.tex}}
\quad+\quad\parbox{20mm}{\input{FeynmanDiagrams/6sQ3.tex}}\quad , \nn
c_{12|3|45|6;\,0}&{}=0\, , \nn
c_{12|3|45|6;\,4}&{}=\frac{2i\left[12\right]\left[54\right]\left[63\right]^{2}}{\left\langle 12\right\rangle \left\langle 45\right\rangle \text{tr}_{5}\left(2,3,6,1\right)\text{tr}_{5}\left(5,3,6,4\right)}\nn
&\qquad\times\left(\left\langle 3\left|1+2\right|3\right]\left\langle 6\left|1+2\right|6\right]-s_{36}s_{12}\right) \, ;  
\end{align}
\end{subequations}

The finite colour-ordered amplitude takes the form
\begin{align}
&A_{6}^{\text{1-loop}}\left(1^{+},2^{+},3^{+},4^{+},5^{+},6^{+}\right)
=c_{123|4|5|6;\,4}\, I_{123|4|5|6}\left[\mu^{4}\right]\nn
&\quad+c_{12|34|5|6;\,4}\, I_{12|34|5|6}\left[\mu^{4}\right]+\frac{1}{2}c_{12|3|45|6;\,4}\, I_{12|3|45|6}\left[\mu^{4}\right]\nn
&\quad+\text{cyclic perms.}
\end{align}
Which agrees with~\cite{Bern:1993qk}.\\

\par
\begin{figure}[htb!]
\begin{subequations}
\begin{align*}
&\parbox{25mm}{\input{FeynmanDiagrams/6gT1.tex}}
\parbox{25mm}{\input{FeynmanDiagrams/6gT2.tex}}
\parbox{25mm}{\input{FeynmanDiagrams/6gT3.tex}}\\
&\parbox{25mm}{\input{FeynmanDiagrams/6gT4.tex}}
\parbox{25mm}{\input{FeynmanDiagrams/6gD1.tex}}
\parbox{25mm}{\input{FeynmanDiagrams/6gD2.tex}}
\end{align*}
\end{subequations}
\caption{Triangle and bubble topologies for the six-point.}\label{6top}
\end{figure}
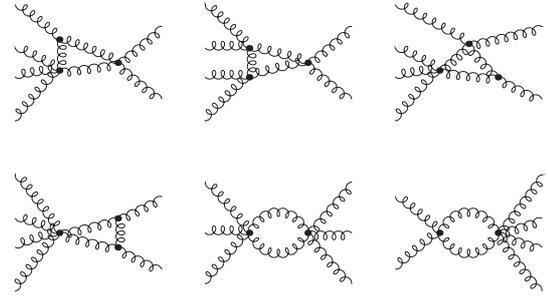

 As done for the five-point, we also consider other helicity configurations, $A_{6}\left(1^{-},2^{+},3^{+},4^{+},5^{+},6^{+}\right)$, $A_{6}\left(1^{-},2^{-},3^{+},4^{+},5^{+},6^{+}\right)$, $A_{6}\left(1^{-},2^{-},3^{-},4^{+},5^{+},6^{+}\right)$, $A_{6}\left(1^{-},2^{+},3^{-},4^{+},5^{-},6^{+}\right)$, $A_{6}\left(1^{-},2^{+},3^{-},4^{+},5^{+},6^{+}\right)$, $A_{6}\left(1^{-},2^{+},3^{+},4^{-},5^{+},6^{+}\right)$ and  $A_{6}\left(1^{-},2^{+},3^{+},4^{-},5^{-},6^{+}\right)$, 
where we also have to consider contributions from triangles and bubbles, such topologies are depicted in fig~\ref{6top}. Moreover, the relevant tree-level are the ones that appear in eq.~(\ref{trees5g}) and the six-points,
\begin{align}
&G\, G\to g\, g\, g\, g,&&S\, S\to g\, g\, g\, g,
\end{align}
where the numerical value of each coefficient in eq.~(\ref{Eq:Decomposition}) agrees with \njet\, for those helicity configurations and with previous results\cite{Mahlon:1993si, Bern:1993qk, Bern:1994zx, Bidder:2004tx, Bedford:2004nh, Britto:2005ha, Bern:2005cq, Bern:2005hh, Britto:2006sj, Berger:2006ci, Berger:2006vq, Xiao:2006vt}.
\vspace{-0.4cm}
\subsection{The gggH Amplitude}
As final example we show the calculation of the leading colour-ordered one-loop helicity  amplitude $A_{6}\left(1^{+},2^{+},3^{+},\text{H}\right)$ in the heavy top mass limit. Since this amplitude is symmetric under cyclic and non-cyclic permutations of the particles we only consider the independent topologies for boxes, triangles and bubbles.
\par The leading-order contribution reads as follows
\begin{align}
A_{4,H}^{\text{tree}}\left(1^{+},2^{+},3^{+},\text{H}\right) & =\frac{-i\, m_{H}^{4}}{\left\langle 12\right\rangle \left\langle 23\right\rangle \left\langle 31\right\rangle },\end{align}
The quadruple cut is given by:
\vspace{-0.4cm}
\begin{subequations}
\begin{align}
C_{1|2|3|H} &{}= 
      \parbox{25mm}{\input{DiagramHiggs/C123H.1.tex}}
+     \parbox{25mm}{\input{DiagramHiggs/C123H.2.tex}}   \, , \nn[-2.ex]
c_{1|2|3|H;\,0}&{}=-\frac{1}{2}A_{4,H}^{\text{tree}}\left(1^{+},2^{+},3^{+},\text{H}\right)s_{12}s_{23}\, , \nn
c_{1|2|3|H;\,4}&{}=0 \, ,
\end{align}
\label{Eq:coeffF}
\end{subequations}

The triple cut with  two massive channels is
\vspace{-0.4cm}
\begin{subequations}
\begin{align}
C_{12|3|H} &{}= 
      \parbox{25mm}{\input{DiagramHiggs/C13H.1.tex}}
+    \parbox{25mm}{\input{DiagramHiggs/C13H.2.tex}}   \, , \nn[-2.ex]
c_{12|3|H;\,0} &{}=\frac{1}{2}A_{4,H}^{\text{tree}}\left(1^{+},2^{+},3^{+},\text{H}\right)\left(s_{13}+s_{23}\right)\, , \nn
c_{12|3|H;\,2} &{}=0  \, ,
\end{align}
\end{subequations}
while the one  with one massive channel only reads as follows:
\vspace{-0.4cm}
\begin{subequations}
\begin{align}
C_{1|2|3H} &{}= 
      \parbox{25mm}{\input{DiagramHiggs/C123a.1.tex}}
+    \parbox{25mm}{\input{DiagramHiggs/C123a.2.tex}}   \, , \nn[-2.ex]
c_{1|2|3H;\,0}&{}=0\, , \nn
c_{1|2|3H;\,0}&{}=-2A_{4,H}^{\text{tree}}\left(1^{+},2^{+},3^{+},\text{H}\right)\frac{s_{13}s_{23}}{m_{H}^{4}}  \, ,
\end{align}
\end{subequations}
Finally, the double cut is given by:
\vspace{-0.4cm}
\begin{subequations}
\begin{align}
C_{12|3H} &{}= 
      \parbox{25mm}{\input{DiagramHiggs/C13.1.tex}}
+    \parbox{25mm}{\input{DiagramHiggs/C13.2.tex}}   \, , \nn[-2.ex]
c_{12|3H;\,0}&{}=0\, , \nn
c_{12|3H;\,2}&{}=4A_{4,H}^{\text{tree}}\left(1^{+},2^{+},3^{+},\text{H}\right)\frac{s_{13}s_{23}}{s_{12}m_{H}^{4}}  \, ,
\end{align}
\label{Eq:coeffL}
\end{subequations}
The cut $C_{123 | H}$ does not give any contribution.
\smallskip
The required trees to compute these coefficients are (\ref{trees5g}) and 
\begin{align}
&G\, G\to H,&&S\, S\to H,\nn
&G\, G\to g\, H,&&S\, S\to g\, H\label{trees4gH}
\end{align}
where the Feynman rules for the Higgs-gluon and Higgs-scalar couplings in the FDF are given in Appendix C of~\cite{Fazio:2014xea}.

\par Then the colour-ordered one-loop helicity amplitude takes the form
\begin{align}
&A_{4,H}^{\text{1-loop}}\left(1^{+},2^{+},3^{+},\text{H}\right)
=c_{1|2|3|H;\,0}\, I_{1|2|3|H;\,0}\nn
&\qquad+c_{12|3|H;\,0}\, I_{12|3|H;\,0}
+c_{1|2|3H;\,2}\, I_{1|2|3H;\,2}\left[\mu^{2}\right]\nn
&\qquad+c_{12|3H;\,2}\, I_{12|3H;\,2}\left[\mu^{2}\right]+\text{cyclic perms.}
\end{align}
Which agrees with~\cite{Schmidt:1997wr}.\\

The procedure for computing the one-loop amplitudes given above has
been fully automated. In particular, we have implemented the FDF Feynman rules (including
the $(-2\epsilon)$-SRs) in \Feynarts/\Feyncalc\,~\cite{Hahn:2000kx}, to build automatically the
tree-level amplitudes to be sewn in the cuts. Then, the coefficients of the master integrals are determined by applying the integrand reduction via Laurent expansion~\cite{Mastrolia:2012bu}, which has been implemented in Mathematica, by using the package \sam~\cite{Maitre:2007jq}.

%% file: DiagramCut/Cg1234.1.tex
\unitlength=0.20bp%
\begin{feynartspicture}(300,300)(1,1)
\FADiagram{}
%
\FAProp(4.,15.)(6.5,13.5)(0.,){/Cycles}{0}
\FAProp(4.,5.)(6.5,6.5)(0.,){/Cycles}{0}
\FAProp(16.,15.)(13.5,13.5)(0.,){/Cycles}{0}
\FAProp(16.,5.)(13.5,6.5)(0.,){/Cycles}{0}
%
\FAProp(6.5,13.5)(6.5,6.5)(0.,){/Cycles}{0}
\FAProp(13.5,13.5)(6.5,13.5)(0.,){/Cycles}{0}
\FAProp(6.5,6.5)(13.5,6.5)(0.,){/Cycles}{0}
\FAProp(13.5,6.5)(13.5,13.5)(0.,){/Cycles}{0}
%
\FAVert(6.5,13.5){0}
\FAVert(6.5,6.5){0}
\FAVert(13.5,13.5){0}
\FAVert(13.5,6.5){0}
%
%
\FALabel(11.1,4.5)[]{\tiny $+$}
\FALabel(8.9,4.5)[]{\tiny $-$}
\FALabel(4.5, 8.9)[]{\tiny $+$}
\FALabel(4.5, 11.1)[]{\tiny $-$}
\FALabel(8.9, 15.)[]{\tiny $+$}
\FALabel(11.1, 15.)[]{\tiny $-$}
\FALabel(15., 11.1)[]{\tiny $+$}
\FALabel(15., 8.9)[]{\tiny $-$}
%
%
\FALabel(3.,4.)[]{\tiny$1^+$}
\FALabel(3.,16.)[]{\tiny$2^+$}
\FALabel(17.,16.)[]{\tiny$3^+$}
\FALabel(17.,4.)[]{\tiny$4^+$}
\end{feynartspicture}

%% file: DiagramCut/Cg1234.2.tex
\unitlength=0.20bp%
\begin{feynartspicture}(300,300)(1,1)
\FADiagram{}
%
\FAProp(4.,15.)(6.5,13.5)(0.,){/Cycles}{0}
\FAProp(4.,5.)(6.5,6.5)(0.,){/Cycles}{0}
\FAProp(16.,15.)(13.5,13.5)(0.,){/Cycles}{0}
\FAProp(16.,5.)(13.5,6.5)(0.,){/Cycles}{0}
%
\FAProp(6.5,13.5)(6.5,6.5)(0.,){/Cycles}{0}
\FAProp(13.5,13.5)(6.5,13.5)(0.,){/Cycles}{0}
\FAProp(6.5,6.5)(13.5,6.5)(0.,){/Cycles}{0}
\FAProp(13.5,6.5)(13.5,13.5)(0.,){/Cycles}{0}
%
\FAVert(6.5,13.5){0}
\FAVert(6.5,6.5){0}
\FAVert(13.5,13.5){0}
\FAVert(13.5,6.5){0}
%
%
\FALabel(11.1,4.5)[]{\tiny $-$}
\FALabel(8.9,4.5)[]{\tiny $+$}
\FALabel(4.5, 8.9)[]{\tiny $-$}
\FALabel(4.5, 11.1)[]{\tiny $+$}
\FALabel(8.9, 15.)[]{\tiny $-$}
\FALabel(11.1, 15.)[]{\tiny $+$}
\FALabel(15., 11.1)[]{\tiny $-$}
\FALabel(15., 8.9)[]{\tiny $+$}
%
%
\FALabel(3.,4.)[]{\tiny$1^+$}
\FALabel(3.,16.)[]{\tiny$2^+$}
\FALabel(17.,16.)[]{\tiny$3^+$}
\FALabel(17.,4.)[]{\tiny$4^+$}
\end{feynartspicture}

%% file: DiagramCut/Cg1234.3.tex
\unitlength=0.20bp%
\begin{feynartspicture}(300,300)(1,1)
\FADiagram{}
%
\FAProp(4.,15.)(6.5,13.5)(0.,){/Cycles}{0}
\FAProp(4.,5.)(6.5,6.5)(0.,){/Cycles}{0}
\FAProp(16.,15.)(13.5,13.5)(0.,){/Cycles}{0}
\FAProp(16.,5.)(13.5,6.5)(0.,){/Cycles}{0}
%
\FAProp(6.5,13.5)(6.5,6.5)(0.,){/Cycles}{0}
\FAProp(13.5,13.5)(6.5,13.5)(0.,){/Cycles}{0}
\FAProp(6.5,6.5)(13.5,6.5)(0.,){/Cycles}{0}
\FAProp(13.5,6.5)(13.5,13.5)(0.,){/Cycles}{0}
%
\FAVert(6.5,13.5){0}
\FAVert(6.5,6.5){0}
\FAVert(13.5,13.5){0}
\FAVert(13.5,6.5){0}
%
%
\FALabel(11.1,4.5)[]{\tiny $0$}
\FALabel(8.9,4.5)[]{\tiny $0$}
\FALabel(4.5, 8.9)[]{\tiny $0$}
\FALabel(4.5, 11.1)[]{\tiny $0$}
\FALabel(8.9, 15.)[]{\tiny $0$}
\FALabel(11.1, 15.)[]{\tiny $0$}
\FALabel(15., 11.1)[]{\tiny $0$}
\FALabel(15., 8.9)[]{\tiny $0$}
%
%
\FALabel(3.,4.)[]{\tiny$1^+$}
\FALabel(3.,16.)[]{\tiny$2^+$}
\FALabel(17.,16.)[]{\tiny$3^+$}
\FALabel(17.,4.)[]{\tiny$4^+$}
\end{feynartspicture}

%% file: DiagramCut/Cs1.tex
\unitlength=0.20bp%
\begin{feynartspicture}(300,300)(1,1)
\FADiagram{}
%
\FAProp(4.,15.)(6.5,13.5)(0.,){/Cycles}{0}
\FAProp(4.,5.)(6.5,6.5)(0.,){/Cycles}{0}
\FAProp(16.,15.)(13.5,13.5)(0.,){/Cycles}{0}
\FAProp(16.,5.)(13.5,6.5)(0.,){/Cycles}{0}
%
\FAProp(6.5,13.5)(6.5,6.5)(0.,){/Cycles}{0}
\FAProp(13.5,13.5)(6.5,13.5)(0.,){/Cycles}{0}
\FAProp(6.5,6.5)(13.5,6.5)(0.,){/Cycles}{0}
\FAProp(13.5,6.5)(13.5,13.5)(0.,){/ScalarDash}{0}
%
\FAVert(6.5,13.5){0}
\FAVert(6.5,6.5){0}
\FAVert(13.5,13.5){0}
\FAVert(13.5,6.5){0}
%
%
\FALabel(10.0,4.5)[]{\tiny $-h_1 h_1$}
\FALabel(4.1, 8.9)[]{\tiny $h_2$}
\FALabel(4.1, 11.1)[]{\tiny $-h_2$}
\FALabel(10.0, 15.)[]{\tiny $-h_3  h_3$}
%
%
\FALabel(3.,4.)[]{\tiny$1^+$}
\FALabel(3.,16.)[]{\tiny$2^+$}
\FALabel(17.,16.)[]{\tiny$3^+$}
\FALabel(17.,4.)[]{\tiny$4^+$}
\end{feynartspicture}

%% file: DiagramCut/Cs2o.tex
\unitlength=0.20bp%
\begin{feynartspicture}(300,300)(1,1)
\FADiagram{}
%
\FAProp(4.,15.)(6.5,13.5)(0.,){/Cycles}{0}
\FAProp(4.,5.)(6.5,6.5)(0.,){/Cycles}{0}
\FAProp(16.,15.)(13.5,13.5)(0.,){/Cycles}{0}
\FAProp(16.,5.)(13.5,6.5)(0.,){/Cycles}{0}
%
\FAProp(6.5,13.5)(6.5,6.5)(0.,){/ScalarDash}{0}
\FAProp(13.5,13.5)(6.5,13.5)(0.,){/Cycles}{0}
\FAProp(6.5,6.5)(13.5,6.5)(0.,){/Cycles}{0}
\FAProp(13.5,6.5)(13.5,13.5)(0.,){/ScalarDash}{0}
%
\FAVert(6.5,13.5){0}
\FAVert(6.5,6.5){0}
\FAVert(13.5,13.5){0}
\FAVert(13.5,6.5){0}
%
%
\FALabel(10.0,4.5)[]{\tiny $-h_1 h_1$}
\FALabel(10.0, 15.)[]{\tiny $-h_2  h_2$}
%
%
\FALabel(3.,4.)[]{\tiny$1^+$}
\FALabel(3.,16.)[]{\tiny$2^+$}
\FALabel(17.,16.)[]{\tiny$3^+$}
\FALabel(17.,4.)[]{\tiny$4^+$}
\end{feynartspicture}

%% file: DiagramCut/Cs2a.tex
\unitlength=0.20bp%
\begin{feynartspicture}(300,300)(1,1)
\FADiagram{}
%
\FAProp(4.,15.)(6.5,13.5)(0.,){/Cycles}{0}
\FAProp(4.,5.)(6.5,6.5)(0.,){/Cycles}{0}
\FAProp(16.,15.)(13.5,13.5)(0.,){/Cycles}{0}
\FAProp(16.,5.)(13.5,6.5)(0.,){/Cycles}{0}
%
\FAProp(6.5,13.5)(6.5,6.5)(0.,){/Cycles}{0}
\FAProp(13.5,13.5)(6.5,13.5)(0.,){/ScalarDash}{0}
\FAProp(6.5,6.5)(13.5,6.5)(0.,){/Cycles}{0}
\FAProp(13.5,6.5)(13.5,13.5)(0.,){/ScalarDash}{0}
%
\FAVert(6.5,13.5){0}
\FAVert(6.5,6.5){0}
\FAVert(13.5,13.5){0}
\FAVert(13.5,6.5){0}
%
%
\FALabel(10.0,4.5)[]{\tiny $-h_1 h_1$}
\FALabel(4.1, 8.9)[]{\tiny $h_2$}
\FALabel(4.1, 11.1)[]{\tiny $-h_2$}
%
%
\FALabel(3.,4.)[]{\tiny$1^+$}
\FALabel(3.,16.)[]{\tiny$2^+$}
\FALabel(17.,16.)[]{\tiny$3^+$}
\FALabel(17.,4.)[]{\tiny$4^+$}
\end{feynartspicture}

%% file: DiagramCut/Cs3.tex
\unitlength=0.20bp%
\begin{feynartspicture}(300,300)(1,1)
\FADiagram{}
%
\FAProp(4.,15.)(6.5,13.5)(0.,){/Cycles}{0}
\FAProp(4.,5.)(6.5,6.5)(0.,){/Cycles}{0}
\FAProp(16.,15.)(13.5,13.5)(0.,){/Cycles}{0}
\FAProp(16.,5.)(13.5,6.5)(0.,){/Cycles}{0}
%
\FAProp(6.5,13.5)(6.5,6.5)(0.,){/ScalarDash}{0}
\FAProp(13.5,13.5)(6.5,13.5)(0.,){/ScalarDash}{0}
\FAProp(6.5,6.5)(13.5,6.5)(0.,){/Cycles}{0}
\FAProp(13.5,6.5)(13.5,13.5)(0.,){/ScalarDash}{0}
%
\FAVert(6.5,13.5){0}
\FAVert(6.5,6.5){0}
\FAVert(13.5,13.5){0}
\FAVert(13.5,6.5){0}
%
%
\FALabel(10.0,4.5)[]{\tiny $-h_1 h_1$}
%
%
\FALabel(3.,4.)[]{\tiny$1^+$}
\FALabel(3.,16.)[]{\tiny$2^+$}
\FALabel(17.,16.)[]{\tiny$3^+$}
\FALabel(17.,4.)[]{\tiny$4^+$}
\end{feynartspicture}

%% file: DiagramCut/Cg1234.4.tex
\unitlength=0.20bp%
\begin{feynartspicture}(300,300)(1,1)
\FADiagram{}
%
\FAProp(4.,15.)(6.5,13.5)(0.,){/Cycles}{0}
\FAProp(4.,5.)(6.5,6.5)(0.,){/Cycles}{0}
\FAProp(16.,15.)(13.5,13.5)(0.,){/Cycles}{0}
\FAProp(16.,5.)(13.5,6.5)(0.,){/Cycles}{0}
%
\FAProp(6.5,13.5)(6.5,6.5)(0.,){/ScalarDash}{0}
\FAProp(13.5,13.5)(6.5,13.5)(0.,){/ScalarDash}{0}
\FAProp(6.5,6.5)(13.5,6.5)(0.,){/ScalarDash}{0}
\FAProp(13.5,13.5)(13.5,6.5)(0.,){/ScalarDash}{0}
%
\FAVert(6.5,13.5){0}
\FAVert(6.5,6.5){0}
\FAVert(13.5,13.5){0}
\FAVert(13.5,6.5){0}
%
%
%
%
\FALabel(3.,4.)[]{\tiny$1^+$}
\FALabel(3.,16.)[]{\tiny$2^+$}
\FALabel(17.,16.)[]{\tiny$3^+$}
\FALabel(17.,4.)[]{\tiny$4^+$}
\end{feynartspicture}

%% file: FeynmanDiagrams/5gQ.tex
\unitlength=0.20bp%
\begin{feynartspicture}(300,300)(1,1)
\FADiagram{}
%
\FAProp(0.,17.)(6.5,13.5)(0.,){/Cycles}{0}
\FAProp(0.,10.)(6.5,6.5)(0.,){/Cycles}{0}
\FAProp(0.,3.)(6.5,6.5)(0.,){/Cycles}{0}
\FAProp(20.,15.)(13.5,13.5)(0.,){/Cycles}{0}
\FAProp(20.,5.)(13.5,6.5)(0.,){/Cycles}{0}
%
\FAProp(6.5,13.5)(13.5,13.5)(0.,){/Cycles}{0}
\FAProp(6.5,13.5)(6.5,6.5)(0.,){/Cycles}{0}
\FAProp(13.5,13.5)(13.5,6.5)(0.,){/Cycles}{0}
\FAProp(13.5,6.5)(6.5,6.5)(0.,){/Cycles}{0}
%
\FAVert(6.5,13.5){0}
\FAVert(13.5,13.5){0}
\FAVert(13.5,6.5){0}
\FAVert(6.5,6.5){0}
%
%
\FALabel(20,5)[l]{$5^+$}
\FALabel(20,15)[l]{$4^+$}
\FALabel(0,17)[r]{$3^+$}
\FALabel(0,10)[r]{$2^+$}
\FALabel(0,3)[r]{$1^+$}
\end{feynartspicture}

%% file: FeynmanDiagrams/5sQ.tex
\unitlength=0.20bp%
\begin{feynartspicture}(300,300)(1,1)
\FADiagram{}
%
\FAProp(0.,17.)(6.5,13.5)(0.,){/Cycles}{0}
\FAProp(0.,10.)(6.5,6.5)(0.,){/Cycles}{0}
\FAProp(0.,3.)(6.5,6.5)(0.,){/Cycles}{0}
\FAProp(20.,15.)(13.5,13.5)(0.,){/Cycles}{0}
\FAProp(20.,5.)(13.5,6.5)(0.,){/Cycles}{0}
\FAProp(6.5,13.5)(13.5,13.5)(0.,){/ScalarDash}{0}
\FAProp(6.5,13.5)(6.5,6.5)(0.,){/ScalarDash}{0}
\FAProp(13.5,13.5)(13.5,6.5)(0.,){/ScalarDash}{0}
\FAProp(13.5,6.5)(6.5,6.5)(0.,){/ScalarDash}{0}
%
\FAVert(6.5,13.5){0}
\FAVert(13.5,13.5){0}
\FAVert(13.5,6.5){0}
\FAVert(6.5,6.5){0}
%
%
\FALabel(20,5)[l]{$5^+$}
\FALabel(20,15)[l]{$4^+$}
\FALabel(0,17)[r]{$3^+$}
\FALabel(0,10)[r]{$2^+$}
\FALabel(0,3)[r]{$1^+$}
\end{feynartspicture}

%% file: FeynmanDiagrams/5gT2.tex
\unitlength=0.20bp%
\begin{feynartspicture}(300,300)(1,1)
\FADiagram{}
%
\FAProp(0.,17.)(10.,12.5)(0.,){/Cycles}{0}
\FAProp(0.,10.)(6.5,6.5)(0.,){/Cycles}{0}
\FAProp(0.,3.)(6.5,6.5)(0.,){/Cycles}{0}
\FAProp(20.,15.)(10.,12.5)(0.,){/Cycles}{0}
\FAProp(20.,5.)(13.5,6.5)(0.,){/Cycles}{0}
\FAProp(13.5,6.5)(10.,12.5)(0.,){/Cycles}{0}
\FAProp(13.5,6.5)(6.5,6.5)(0.,){/Cycles}{0}
\FAProp(10.,12.5)(6.5,6.5)(0.,){/Cycles}{0}
%
\FAVert(13.5,6.5){0}
\FAVert(10.,12.5){0}
\FAVert(6.5,6.5){0}
%
%
%
\end{feynartspicture}

%% file: FeynmanDiagrams/5gT3.tex
\unitlength=0.20bp%
\begin{feynartspicture}(300,300)(1,1)
\FADiagram{}
%
\FAProp(0.,17.)(6.,10.)(0.,){/Cycles}{0}
\FAProp(0.,10.)(6.,10.)(0.,){/Cycles}{0}
\FAProp(0.,3.)(6.,10.)(0.,){/Cycles}{0}
\FAProp(20.,15.)(14.,12.)(0.,){/Cycles}{0}
\FAProp(20.,5.)(14.,8.)(0.,){/Cycles}{0}
\FAProp(14.,12.)(14.,8.)(0.,){/Cycles}{0}
\FAProp(14.,12.)(6.,10.)(0.,){/Cycles}{0}
\FAProp(14.,8.)(6.,10.)(0.,){/Cycles}{0}
%
\FAVert(14.,12.){0}
\FAVert(14.,8.){0}
\FAVert(6.,10.){0}
%
%
%
\end{feynartspicture}

%% file: FeynmanDiagrams/5gD.tex
\unitlength=0.20bp%
\begin{feynartspicture}(300,300)(1,1)
\FADiagram{}
%
\FAProp(0.,17.)(6.,10.)(0.,){/Cycles}{0}
\FAProp(0.,10.)(6.,10.)(0.,){/Cycles}{0}
\FAProp(0.,3.)(6.,10.)(0.,){/Cycles}{0}
\FAProp(20.,15.)(14.,10.)(0.,){/Cycles}{0}
\FAProp(20.,5.)(14.,10.)(0.,){/Cycles}{0}
\FAProp(14.,10.)(6.,10.)(0.8,){/Cycles}{0}
\FAProp(14.,10.)(6.,10.)(-0.8,){/Cycles}{0}
%
\FAVert(14.,10.){0}
\FAVert(6.,10.){0}
%
%
%
\end{feynartspicture}

%% file: FeynmanDiagrams/6gQ1.tex
\unitlength=0.20bp%
\begin{feynartspicture}(300,300)(1,1)
\FADiagram{}
%
\FAProp(0.,18.)(6.,13.2)(0.,){/Cycles}{0}
\FAProp(0.,12.)(6.,8.93333)(0.,){/Cycles}{0}
\FAProp(0.,8.)(6.,8.93333)(0.,){/Cycles}{0}
\FAProp(0.,2.)(6.,8.93333)(0.,){/Cycles}{0}
\FAProp(20.,15.)(14.,12.)(0.,){/Cycles}{0}
\FAProp(20.,5.)(14.,8.)(0.,){/Cycles}{0}
%
\FAProp(6.,13.2)(14.,12.)(0.,){/Cycles}{0}
\FAProp(6.,13.2)(6.,8.93333)(0.,){/Cycles}{0}
\FAProp(14.,12.)(14.,8.)(0.,){/Cycles}{0}
\FAProp(14.,8.)(6.,8.93333)(0.,){/Cycles}{0}
%
\FAVert(6.,13.2){0}
\FAVert(14.,12.){0}
\FAVert(14.,8.){0}
\FAVert(6.,8.93333){0}
%
%
\FALabel(0,18)[r]{$4^+$}
\FALabel(0,12)[r]{$3^+$}
\FALabel(0,8)[r]{$2^+$}
\FALabel(0,2)[r]{$1^+$}
\FALabel(20,15)[l]{$5^+$}
\FALabel(20,5)[l]{$6^+$}
\end{feynartspicture}

%% file: FeynmanDiagrams/6sQ1.tex
\unitlength=0.20bp%
\begin{feynartspicture}(300,300)(1,1)
\FADiagram{}
%
\FAProp(0.,18.)(6.,13.2)(0.,){/Cycles}{0}
\FAProp(0.,12.)(6.,8.93333)(0.,){/Cycles}{0}
\FAProp(0.,8.)(6.,8.93333)(0.,){/Cycles}{0}
\FAProp(0.,2.)(6.,8.93333)(0.,){/Cycles}{0}
\FAProp(20.,15.)(14.,12.)(0.,){/Cycles}{0}
\FAProp(20.,5.)(14.,8.)(0.,){/Cycles}{0}
%
\FAProp(6.,13.2)(14.,12.)(0.,){/ScalarDash}{0}
\FAProp(6.,13.2)(6.,8.93333)(0.,){/ScalarDash}{0}
\FAProp(14.,12.)(14.,8.)(0.,){/ScalarDash}{0}
\FAProp(14.,8.)(6.,8.93333)(0.,){/ScalarDash}{0}
%
\FAVert(6.,13.2){0}
\FAVert(14.,12.){0}
\FAVert(14.,8.){0}
\FAVert(6.,8.93333){0}
%
%
\FALabel(0,18)[r]{$4^+$}
\FALabel(0,12)[r]{$3^+$}
\FALabel(0,8)[r]{$2^+$}
\FALabel(0,2)[r]{$1^+$}
\FALabel(20,15)[l]{$5^+$}
\FALabel(20,5)[l]{$6^+$}
\end{feynartspicture}

%% file: FeynmanDiagrams/6gQ2.tex
\unitlength=0.20bp%
\begin{feynartspicture}(300,300)(1,1)
\FADiagram{}
%
\FAProp(0.,18.)(6.,12.)(0.,){/Cycles}{0}
\FAProp(0.,12.)(6.,12.)(0.,){/Cycles}{0}
\FAProp(0.,8.)(6.,8.)(0.,){/Cycles}{0}
\FAProp(0.,2.)(6.,8.)(0.,){/Cycles}{0}
\FAProp(20.,15.)(14.,12.)(0.,){/Cycles}{0}
\FAProp(20.,5.)(14.,8.)(0.,){/Cycles}{0}
%
\FAProp(14.,12.)(14.,8.)(0.,){/Cycles}{0}
\FAProp(14.,12.)(6.,12.)(0.,){/Cycles}{0}
\FAProp(14.,8.)(6.,8.)(0.,){/Cycles}{0}
\FAProp(6.,12.)(6.,8.)(0.,){/Cycles}{0}
%
\FAVert(14.,12.){0}
\FAVert(14.,8.){0}
\FAVert(6.,12.){0}
\FAVert(6.,8.){0}
%
%
\FALabel(0,18)[r]{$4^+$}
\FALabel(0,12)[r]{$3^+$}
\FALabel(0,8)[r]{$2^+$}
\FALabel(0,2)[r]{$1^+$}
\FALabel(20,15)[l]{$5^+$}
\FALabel(20,5)[l]{$6^+$}
\end{feynartspicture}

%% file: FeynmanDiagrams/6sQ2.tex
\unitlength=0.20bp%
\begin{feynartspicture}(300,300)(1,1)
\FADiagram{}
%
\FAProp(0.,18.)(6.,12.)(0.,){/Cycles}{0}
\FAProp(0.,12.)(6.,12.)(0.,){/Cycles}{0}
\FAProp(0.,8.)(6.,8.)(0.,){/Cycles}{0}
\FAProp(0.,2.)(6.,8.)(0.,){/Cycles}{0}
\FAProp(20.,15.)(14.,12.)(0.,){/Cycles}{0}
\FAProp(20.,5.)(14.,8.)(0.,){/Cycles}{0}
%
\FAProp(14.,12.)(14.,8.)(0.,){/ScalarDash}{0}
\FAProp(14.,12.)(6.,12.)(0.,){/ScalarDash}{0}
\FAProp(14.,8.)(6.,8.)(0.,){/ScalarDash}{0}
\FAProp(6.,12.)(6.,8.)(0.,){/ScalarDash}{0}
%
\FAVert(14.,12.){0}
\FAVert(14.,8.){0}
\FAVert(6.,12.){0}
\FAVert(6.,8.){0}
%
%
\FALabel(0,18)[r]{$4^+$}
\FALabel(0,12)[r]{$3^+$}
\FALabel(0,8)[r]{$2^+$}
\FALabel(0,2)[r]{$1^+$}
\FALabel(20,15)[l]{$5^+$}
\FALabel(20,5)[l]{$6^+$}
\end{feynartspicture}

%% file: FeynmanDiagrams/6gQ3.tex
\unitlength=0.20bp%
\begin{feynartspicture}(300,300)(1,1)
\FADiagram{}
%
\FAProp(0.,17.)(6.,12.8)(0.,){/Cycles}{0}
\FAProp(0.,10.)(6.,8.6)(0.,){/Cycles}{0}
\FAProp(0.,3.)(6.,8.6)(0.,){/Cycles}{0}
\FAProp(20.,17.)(14.,11.4)(0.,){/Cycles}{0}
\FAProp(20.,10.)(14.,11.4)(0.,){/Cycles}{0}
\FAProp(20.,3.)(14.,7.2)(0.,){/Cycles}{0}
%
\FAProp(6.,12.8)(6.,8.6)(0.,){/Cycles}{0}
\FAProp(6.,12.8)(14.,11.4)(0.,){/Cycles}{0}
\FAProp(14.,7.2)(6.,8.6)(0.,){/Cycles}{0}
\FAProp(14.,7.2)(14.,11.4)(0.,){/Cycles}{0}
%
\FAVert(6.,12.8){0}
\FAVert(14.,7.2){0}
\FAVert(6.,8.6){0}
\FAVert(14.,11.4){0}
%
%
\FALabel(0,17)[r]{$3^+$}
\FALabel(0,10)[r]{$2^+$}
\FALabel(0,3)[r]{$1^+$}
\FALabel(20,17)[l]{$4^+$}
\FALabel(20,10)[l]{$5^+$}
\FALabel(20,3)[l]{$6^+$}
\end{feynartspicture}

%% file: FeynmanDiagrams/6sQ3.tex
\unitlength=0.20bp%
\begin{feynartspicture}(300,300)(1,1)
\FADiagram{}
%
\FAProp(0.,17.)(6.,12.8)(0.,){/Cycles}{0}
\FAProp(0.,10.)(6.,8.6)(0.,){/Cycles}{0}
\FAProp(0.,3.)(6.,8.6)(0.,){/Cycles}{0}
\FAProp(20.,17.)(14.,11.4)(0.,){/Cycles}{0}
\FAProp(20.,10.)(14.,11.4)(0.,){/Cycles}{0}
\FAProp(20.,3.)(14.,7.2)(0.,){/Cycles}{0}
%
\FAProp(6.,12.8)(6.,8.6)(0.,){/ScalarDash}{0}
\FAProp(6.,12.8)(14.,11.4)(0.,){/ScalarDash}{0}
\FAProp(14.,7.2)(6.,8.6)(0.,){/ScalarDash}{0}
\FAProp(14.,7.2)(14.,11.4)(0.,){/ScalarDash}{0}
%
\FAVert(6.,12.8){0}
\FAVert(14.,7.2){0}
\FAVert(6.,8.6){0}
\FAVert(14.,11.4){0}
%
%
\FALabel(0,17)[r]{$3^+$}
\FALabel(0,10)[r]{$2^+$}
\FALabel(0,3)[r]{$1^+$}
\FALabel(20,17)[l]{$4^+$}
\FALabel(20,10)[l]{$5^+$}
\FALabel(20,3)[l]{$6^+$}
\end{feynartspicture}

%% file: FeynmanDiagrams/6gT1.tex
\unitlength=0.20bp%
\begin{feynartspicture}(300,300)(1,1)
\FADiagram{}
%
\FAProp(0.,18.)(6.,13.2)(0.,){/Cycles}{0}
\FAProp(0.,12.)(6.,8.93333)(0.,){/Cycles}{0}
\FAProp(0.,8.)(6.,8.93333)(0.,){/Cycles}{0}
\FAProp(0.,2.)(6.,8.93333)(0.,){/Cycles}{0}
\FAProp(20.,15.)(14.,10.)(0.,){/Cycles}{0}
\FAProp(20.,5.)(14.,10.)(0.,){/Cycles}{0}
\FAProp(6.,13.2)(14.,10.)(0.,){/Cycles}{0}
\FAProp(6.,13.2)(6.,8.93333)(0.,){/Cycles}{0}
\FAProp(14.,10.)(6.,8.93333)(0.,){/Cycles}{0}
%
\FAVert(6.,13.2){0}
\FAVert(14.,10.){0}
\FAVert(6.,8.93333){0}
%
%
%
\end{feynartspicture}

%% file: FeynmanDiagrams/6gT2.tex
\unitlength=0.20bp%
\begin{feynartspicture}(300,300)(1,1)
\FADiagram{}
%
\FAProp(0.,18.)(6.,12.)(0.,){/Cycles}{0}
\FAProp(0.,12.)(6.,12.)(0.,){/Cycles}{0}
\FAProp(0.,8.)(6.,8.)(0.,){/Cycles}{0}
\FAProp(0.,2.)(6.,8.)(0.,){/Cycles}{0}
\FAProp(20.,15.)(14.,10.)(0.,){/Cycles}{0}
\FAProp(20.,5.)(14.,10.)(0.,){/Cycles}{0}
\FAProp(6.,12.)(6.,8.)(0.,){/Cycles}{0}
\FAProp(6.,12.)(14.,10.)(0.,){/Cycles}{0}
\FAProp(6.,8.)(14.,10.)(0.,){/Cycles}{0}
%
\FAVert(6.,12.){0}
\FAVert(6.,8.){0}
\FAVert(14.,10.){0}
%
%
%
\end{feynartspicture}

%% file: FeynmanDiagrams/6gT3.tex
\unitlength=0.20bp%
\begin{feynartspicture}(300,300)(1,1)
\FADiagram{}
%
\FAProp(0.,18.)(10.,12.6)(0.,){/Cycles}{0}
\FAProp(0.,12.)(6.,8.93333)(0.,){/Cycles}{0}
\FAProp(0.,8.)(6.,8.93333)(0.,){/Cycles}{0}
\FAProp(0.,2.)(6.,8.93333)(0.,){/Cycles}{0}
\FAProp(20.,15.)(10.,12.6)(0.,){/Cycles}{0}
\FAProp(20.,5.)(14.,8.)(0.,){/Cycles}{0}
\FAProp(14.,8.)(10.,12.6)(0.,){/Cycles}{0}
\FAProp(14.,8.)(6.,8.93333)(0.,){/Cycles}{0}
\FAProp(10.,12.6)(6.,8.93333)(0.,){/Cycles}{0}
%
\FAVert(14.,8.){0}
\FAVert(10.,12.6){0}
\FAVert(6.,8.93333){0}
%
%
%
\end{feynartspicture}

%% file: FeynmanDiagrams/6gT4.tex
\unitlength=0.20bp%
\begin{feynartspicture}(300,300)(1,1)
\FADiagram{}
%
\FAProp(0.,18.)(6.,10.)(0.,){/Cycles}{0}
\FAProp(0.,12.)(6.,10.)(0.,){/Cycles}{0}
\FAProp(0.,8.)(6.,10.)(0.,){/Cycles}{0}
\FAProp(0.,2.)(6.,10.)(0.,){/Cycles}{0}
\FAProp(20.,15.)(14.,12.)(0.,){/Cycles}{0}
\FAProp(20.,5.)(14.,8.)(0.,){/Cycles}{0}
\FAProp(14.,12.)(14.,8.)(0.,){/Cycles}{0}
\FAProp(14.,12.)(6.,10.)(0.,){/Cycles}{0}
\FAProp(14.,8.)(6.,10.)(0.,){/Cycles}{0}
%
\FAVert(14.,12.){0}
\FAVert(14.,8.){0}
\FAVert(6.,10.){0}
%
%
%
\end{feynartspicture}

%% file: FeynmanDiagrams/6gD1.tex
\unitlength=0.20bp%
\begin{feynartspicture}(300,300)(1,1)
\FADiagram{}
%
\FAProp(0.,17.)(6.,10.)(0.,){/Cycles}{0}
\FAProp(0.,10.)(6.,10.)(0.,){/Cycles}{0}
\FAProp(0.,3.)(6.,10.)(0.,){/Cycles}{0}
\FAProp(20.,17.)(14.,10.)(0.,){/Cycles}{0}
\FAProp(20.,10.)(14.,10.)(0.,){/Cycles}{0}
\FAProp(20.,3.)(14.,10.)(0.,){/Cycles}{0}
\FAProp(6.,10.)(14.,10.)(0.8,){/Cycles}{0}
\FAProp(6.,10.)(14.,10.)(-0.8,){/Cycles}{0}
%
\FAVert(6.,10.){0}
\FAVert(14.,10.){0}
%
%
%
\end{feynartspicture}

%% file: FeynmanDiagrams/6gD2.tex
\unitlength=0.20bp%
\begin{feynartspicture}(300,300)(1,1)
\FADiagram{}
%
\FAProp(0.,15.)(6.,10.)(0.,){/Cycles}{0}
\FAProp(0.,5.)(6.,10.)(0.,){/Cycles}{0}
\FAProp(20.,18.)(14.,10.)(0.,){/Cycles}{0}
\FAProp(20.,12.)(14.,10.)(0.,){/Cycles}{0}
\FAProp(20.,8.)(14.,10.)(0.,){/Cycles}{0}
\FAProp(20.,2.)(14.,10.)(0.,){/Cycles}{0}
\FAProp(6.,10.)(14.,10.)(-0.8,){/Cycles}{0}
\FAProp(6.,10.)(14.,10.)(0.8,){/Cycles}{0}
%
\FAVert(6.,10.){0}
\FAVert(14.,10.){0}
%
%
%
\end{feynartspicture}

%% file: DiagramHiggs/C123H.1.tex
\unitlength=0.25bp%
\begin{feynartspicture}(300,300)(1,1)
\FADiagram{}
%
\FAProp(4.,15.)(6.5,13.5)(0.,){/Cycles}{0}
\FAProp(4.,5.)(6.5,6.5)(0.,){/Cycles}{0}
\FAProp(16.,15.)(13.5,13.5)(0.,){/Cycles}{0}
\FAProp(16.,5.)(13.5,6.5)(0.,){/ScalarDash}{0}
%
\FAProp(6.5,13.5)(6.5,6.5)(0.,){/Cycles}{0}
\FAProp(13.5,13.5)(6.5,13.5)(0.,){/Cycles}{0}
\FAProp(6.5,6.5)(13.5,6.5)(0.,){/Cycles}{0}
\FAProp(13.5,13.5)(13.5,6.5)(0.,){/Cycles}{0}
%
\FAVert(6.5,13.5){0}
\FAVert(6.5,6.5){0}
\FAVert(13.5,13.5){0}
\FAVert(13.5,6.5){0}
%
%
%
%
\FALabel(3.,4.)[]{\tiny$1^+$}
\FALabel(3.,16.)[]{\tiny$2^+$}
\FALabel(17.,16.)[]{\tiny$3^+$}
\FALabel(17.,4.)[]{\tiny$H$}
\end{feynartspicture}

%% file: DiagramHiggs/C123H.2.tex
\unitlength=0.25bp%
\begin{feynartspicture}(300,300)(1,1)
\FADiagram{}
%
\FAProp(4.,15.)(6.5,13.5)(0.,){/Cycles}{0}
\FAProp(4.,5.)(6.5,6.5)(0.,){/Cycles}{0}
\FAProp(16.,15.)(13.5,13.5)(0.,){/Cycles}{0}
\FAProp(16.,5.)(13.5,6.5)(0.,){/ScalarDash}{0}
%
\FAProp(6.5,13.5)(6.5,6.5)(0.,){/ScalarDash}{0}
\FAProp(13.5,13.5)(6.5,13.5)(0.,){/ScalarDash}{0}
\FAProp(6.5,6.5)(13.5,6.5)(0.,){/ScalarDash}{0}
\FAProp(13.5,13.5)(13.5,6.5)(0.,){/ScalarDash}{0}
%
\FAVert(6.5,13.5){0}
\FAVert(6.5,6.5){0}
\FAVert(13.5,13.5){0}
\FAVert(13.5,6.5){0}
%
%
%
%
\FALabel(3.,4.)[]{\tiny$1^+$}
\FALabel(3.,16.)[]{\tiny$2^+$}
\FALabel(17.,16.)[]{\tiny$3^+$}
\FALabel(17.,4.)[]{\tiny$H$}
\end{feynartspicture}

%% file: DiagramHiggs/C13H.1.tex
\unitlength=0.25bp%
\begin{feynartspicture}(300,300)(1,1)
\FADiagram{}
%
\FAProp(4.,11.5)(6.5,10.0)(0.,){/Cycles}{0}
\FAProp(6.5,10.0)(4.,8.5)(0.,){/Cycles}{0}
\FAProp(16.,15.)(13.5,13.5)(0.,){/Cycles}{0}
\FAProp(16.,5.)(13.5,6.5)(0.,){/ScalarDash}{0}
%
\FAProp(13.5,13.5)(6.5,10.0)(0.,){/Cycles}{0}
\FAProp(6.5,10.0)(13.5,6.5)(0.,){/Cycles}{0}
\FAProp(13.5,13.5)(13.5,6.5)(0.,){/Cycles}{0}
%
%
\FAVert(13.5,13.5){0}
\FAVert(13.5,6.5){0}
\FAVert(6.5,10.){0}
%
%
%
%
\FALabel(3.1,8.)[]{\tiny$1^+$}
\FALabel(3.1,12.)[]{\tiny$2^+$}
\FALabel(17.,16.)[]{\tiny$3^+$}
\FALabel(17.,4.)[]{\tiny$H$}
\end{feynartspicture}

%% file: DiagramHiggs/C13H.2.tex
\unitlength=0.25bp%
\begin{feynartspicture}(300,300)(1,1)
\FADiagram{}
%
\FAProp(4.,11.5)(6.5,10.0)(0.,){/Cycles}{0}
\FAProp(6.5,10.0)(4.,8.5)(0.,){/Cycles}{0}
\FAProp(16.,15.)(13.5,13.5)(0.,){/Cycles}{0}
\FAProp(16.,5.)(13.5,6.5)(0.,){/ScalarDash}{0}
%
\FAProp(13.5,13.5)(6.5,10.0)(0.,){/ScalarDash}{0}
\FAProp(6.5,10.0)(13.5,6.5)(0.,){/ScalarDash}{0}
\FAProp(13.5,13.5)(13.5,6.5)(0.,){/ScalarDash}{0}
%
%
\FAVert(13.5,13.5){0}
\FAVert(13.5,6.5){0}
\FAVert(6.5,10.){0}
%
%
%
%
\FALabel(3.1,8.)[]{\tiny$1^+$}
\FALabel(3.1,12.)[]{\tiny$2^+$}
\FALabel(17.,16.)[]{\tiny$3^+$}
\FALabel(17.,4.)[]{\tiny$H$}
\end{feynartspicture}

%% file: DiagramHiggs/C123a.1.tex
\unitlength=0.25bp%
\begin{feynartspicture}(300,300)(1,1)
\FADiagram{}
%
\FAProp(4.,15.)(6.5,13.5)(0.,){/Cycles}{0}
\FAProp(4.,5.)(6.5,6.5)(0.,){/Cycles}{0}
\FAProp(16.,11.5)(13.5,10.0)(0.,){/Cycles}{0}
\FAProp(16.,8.5)(13.5,10.0)(0.,){/ScalarDash}{0}
%
\FAProp(6.5,13.5)(6.5,6.5)(0.,){/Cycles}{0}
\FAProp(13.5,10.0)(6.5,13.5)(0.,){/Cycles}{0}
\FAProp(6.5,6.5)(13.5,10.0)(0.,){/Cycles}{0}
%
\FAVert(6.5,13.5){0}
\FAVert(6.5,6.5){0}
\FAVert(13.5,10.0){0}
%
%
%
%
\FALabel(3.,4.)[]{\tiny$1^-$}
\FALabel(3.,16.)[]{\tiny$2^+$}
\FALabel(17.,12.)[]{\tiny$3^+$}
\FALabel(17.,8.)[]{\tiny$H$}
\end{feynartspicture}

%% file: DiagramHiggs/C123a.2.tex
\unitlength=0.25bp%
\begin{feynartspicture}(300,300)(1,1)
\FADiagram{}
%
\FAProp(4.,15.)(6.5,13.5)(0.,){/Cycles}{0}
\FAProp(4.,5.)(6.5,6.5)(0.,){/Cycles}{0}
\FAProp(16.,11.5)(13.5,10.0)(0.,){/Cycles}{0}
\FAProp(16.,8.5)(13.5,10.0)(0.,){/ScalarDash}{0}
%
\FAProp(6.5,13.5)(6.5,6.5)(0.,){/ScalarDash}{0}
\FAProp(13.5,10.0)(6.5,13.5)(0.,){/ScalarDash}{0}
\FAProp(6.5,6.5)(13.5,10.0)(0.,){/ScalarDash}{0}
%
\FAVert(6.5,13.5){0}
\FAVert(6.5,6.5){0}
\FAVert(13.5,10.0){0}
%
%
%
%
\FALabel(3.,4.)[]{\tiny$1^-$}
\FALabel(3.,16.)[]{\tiny$2^+$}
\FALabel(17.,12.)[]{\tiny$3^+$}
\FALabel(17.,8.)[]{\tiny$H$}
\end{feynartspicture}

%% file: DiagramHiggs/C13.1.tex
\unitlength=0.25bp%
\begin{feynartspicture}(300,300)(1,1)
\FADiagram{}
\FAProp(4.,11.5)(6.5,10.0)(0.,){/Cycles}{0}
\FAProp(6.5,10.0)(4.,8.5)(0.,){/Cycles}{0}
\FAProp(16.,11.5)(13.5,10.0)(0.,){/Cycles}{0}
\FAProp(16.,8.5)(13.5,10.0)(0.,){/ScalarDash}{0}
\FAVert(6.5,10.){0}
\FAVert(13.5,10.0){0}
\FALabel(3.1,8.)[]{\tiny$1^+$}
\FALabel(3.1,12.)[]{\tiny$2^+$}
\FALabel(17.,12.)[]{\tiny$3^+$}
\FALabel(17.,8.)[]{\tiny$H$}
\FAProp(6.5,10.)(13.5,10.0)(0.8,){/Cycles}{0}
\FAProp(6.5,10.)(13.5,10.0)(-0.8,){/Cycles}{0}
%
\end{feynartspicture}

%% file: DiagramHiggs/C13.2.tex
\unitlength=0.25bp%
\begin{feynartspicture}(300,300)(1,1)
\FADiagram{}
\FAProp(4.,11.5)(6.5,10.0)(0.,){/Cycles}{0}
\FAProp(6.5,10.0)(4.,8.5)(0.,){/Cycles}{0}
\FAProp(16.,11.5)(13.5,10.0)(0.,){/Cycles}{0}
\FAProp(16.,8.5)(13.5,10.0)(0.,){/ScalarDash}{0}
\FAVert(6.5,10.){0}
\FAVert(13.5,10.0){0}
\FALabel(3.1,8.)[]{\tiny$1^+$}
\FALabel(3.1,12.)[]{\tiny$2^+$}
\FALabel(17.,12.)[]{\tiny$3^+$}
\FALabel(17.,8.)[]{\tiny$H$}
\FAProp(6.5,10.)(13.5,10.0)(0.8,){/ScalarDash}{0}
\FAProp(6.5,10.)(13.5,10.0)(-0.8,){/ScalarDash}{0}
%
\end{feynartspicture}

%% file: Section6.tex
\section{Conclusions}
At one-loop level, we have explored the unitarity methods and we have provided a new formalism with extended helicity spinors and consequently extended polarisation vectors, which allows for fully reconstructing the full one loop scattering amplitude. We remark that there is an unified formalism in which the cut-constructible part and the rational part of a scattering amplitude can be found at once. It is enough just to give off-shellness to the internal momentum in a natural way related to the dimensional regularisation and then to perform multiple unitarity cuts for massive internal legs, where for a massless theory such a mass is exactly the off-shellness.

We have presented a set of very non-trivial examples, showing that FDF scheme is suitable for computing important $2 \to 2,3,4$ partonic amplitudes at the next-to-leading order.

\par \bigskip There are many outlooks for this job; they involve the computation of the analytical expressions for $\text{Higgs} +2(3)$ jets in the final state, and also  the two-loop implementation of our formalism. 